\documentclass[preprint2]{aastex}
\shorttitle{Distances to Two SNRs} \shortauthors{Xu et al.}

\begin{document}

\title{Distances to Two Galactic Supernova Remnants: \\
    G32.8-0.1 and G346.6-0.2}

\author{Jian-Wen Xu\altaffilmark{1} and Hui-Rong Zhang\altaffilmark{2}}
\affil{Key Laboratory of Frontiers in Theoretical Physics, Institute
of Theoretical Physics, Chinese Academy of Sciences, Beijing 100080,
China}

\altaffiltext{1}{Postdoctor, Institute of Theoretical Physics,
Chinese Academy of Sciences, Beijing 100080, China.}
\altaffiltext{2}{Postdoctor, Institute of Theoretical Physics,
Chinese Academy of Sciences, Beijing 100080, China.}
\email{xjw@itp.ac.cn}

\begin{abstract}
There are either a near kinematic distance of 5.5~kpc or a far
distance of 8.8~kpc for a Galactic supernova remnant (SNR)
G32.8$-$0.1 derived by using the rotation curve of the Galaxy. Here
we make sure that the remnant distance is the farther one 8.8~kpc
through solving a group of equations for the shell-type remnants
separately at the adiabatic-phase and the radiative-phase. For SNR
G346.6$-$0.2 we determine its distance also the farther one 11~kpc
rather than the nearer one 5.5~kpc.
\end{abstract}


\section{Introduction}

\begin{figure}[htp!!]
\epsscale{1.1} \plotone{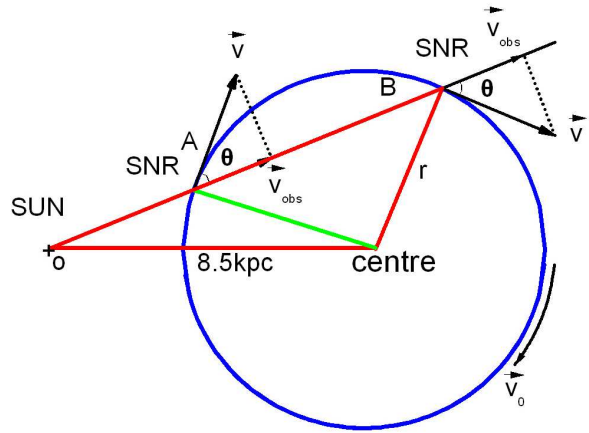}
\caption{Plot illuminates how a near distance OA and a far distance
OB to a calibrator SNR is derived. In this case the remnant distance
to the Galactic center should be less than 8.5~kpc, but cannot be
too near to the center, i.e. $0^o<l<90^o$ or $270^o<l< 360^o$. Here
$l$ is the galactic longitude.}\label{distance2}
\end{figure}

Distances to SNRs can be estimated by observations of extinction,
$X$-ray, SN magnitude, background object, SNR kinematics and HI
absorption, etc. \citep{str88}. In some literatures, the relation
between the radio surface brightness ($\Sigma$) and the linear
diameter ($D$) is also used to determine the distance when the SNR
flux is available \citep{pov68,cla76,loz81,hua85,dur86,gus03}.
However, for the remnants inside the circle of the Galaxy plane with
its radius ($R$) less than 8.5~kpc yet not too near to the Galactic
center, that is the the galactic longitude ($l$) should be
$0^o<l<90^o$ or $270^o<l< 360^o$, then the rotation curve of the
Galaxy can be used to derive the SNR distance after measuring and
obtaining its LSR velocity (Fig.~\ref{distance2}). But usually this
method may lead to two distance values of a near one OA and another
farther one OB. For two examples here, \citet{kor98} derived the
kinematic distance to a shell-type remnant SNR G32.8$-$0.1 either a
near distance of 5.5~kpc or a far distance of 8.8~kpc, and to
another shell-type remnant SNR G346.6$-$0.2 yields a near value of
5.5~kpc and a far value of 11~kpc.

Through a group of equations for the shell-type SNRs separately at
the adiabatic-phase or the radiative-phase, we can determine their
distances when the SN initial explosion energy ($E_0$), the radio
flux at 1~GHz and the observational angle ($\theta$) have been
detected. The relation between the surface brightness ($\Sigma$) and
the remnant diameter ($D$) can also be used to confirm its distance.
But this method is not adopted here because of its somewhat large
deviation.

On the paper, we do numerical calculations of the group of equations
at both stages individually in Sect.~\ref{calculate}, and make some
discussion in Sect.~\ref{discuss}. At last we summarize our
conclusion.

\section{Numerical calculations}\label{calculate}

\subsection{At adiabatic phase}\label{sedov}

Let us list the following group of equations for shell-type remnants
at the second stage \citep{wan84,koy87,big88,xu05},

\begin{equation}
D_{pc} = 4.3\times 10^{-11} \left( \begin{array}{c}
\frac{E_0}{n_{cm^{-3}}} \end{array} \right) ^{1/5} t_{yr} ^{2/5}
\end{equation}

\begin{eqnarray}
\Sigma(D) = 1.505\times 10^{-19} \frac{S_{1GHz}}{\theta_{arcmin}^2}\nonumber\\
= 2.88\times 10^{-14} D_{pc}^{-3.8} n_{cm^{-3}}^2
\end{eqnarray}

\begin{eqnarray}
\left( \begin{array}{c} \frac{E_0}{10^{51}ergs} \end{array} \right)
= 5.3\times10^{-7} n_{cm^{-3}}^{1.12} \upsilon _{Km~s^{-1}}^{1.4}\nonumber\\
\times \left(\begin{array}{c} \frac{D_{pc}}{2}
\end{array} \right)^{3.12}
\end{eqnarray}

Here, $D_{pc}$ is the SNR diameter in units of pc, $t_{yr}$ is the
remnant age in year, $n$ is the ISM electron density in $cm^{-3}$,
$S_{1GHz}$ is the detected fluxes of an SNR in Jy at 1~GHz,
$\theta_{arcmin}$ is the viewing angle in arcmin, $\upsilon =
\frac{dD}{dt}$ is the velocity of shock waves in km~$s^{-1}$. And we
know $tan \left( \begin{array}{c} \frac{\theta_{arcmin}}{2}
\end{array} \right) = \frac{D_{pc}}{2d_{pc}}$, where, $d_{pc}$ is
the distance to a remnant in pc.

\begin{table}
\begin{center}
\caption{Derived distances and diameters of SNR G32.8$-$0.1 and
G346.6$-$0.2 at Sedov-phase after assuming several different initial
explosion energies.\label{tabsedov}}
\begin{tabular}{rrrrr}
\tableline \tableline
 & \multicolumn{2}{c}{G32.8$-$0.1}
 & \multicolumn{2}{c}{G346.6$-$0.2}\\
\tableline
 $E_0$~(ergs) & d(kpc) & D(pc) & d(kpc) & D(pc)\\
\tableline
 $10^{52}$ & 10 & 50 & 19 & 44\\
 $5\times 10^{51}$ & 9.1 & 45 & 17 & 39\\
 $\bf{10^{51}}$ & $\bf{7.0}$ & 35 & $\bf{13}$ & 30\\
 $5\times 10^{50}$ & 6.2 & 31 & 11.5 & 27\\
 $10^{50}$ & 4.8 & 24 & 8.8 & 21\\
 $5\times 10^{49}$ & 4.3 & 21 & 7.9 & 18\\
\tableline
\end{tabular}
\end{center}
\end{table}

\begin{figure}[htp!!]
\epsscale{1.1} \plotone{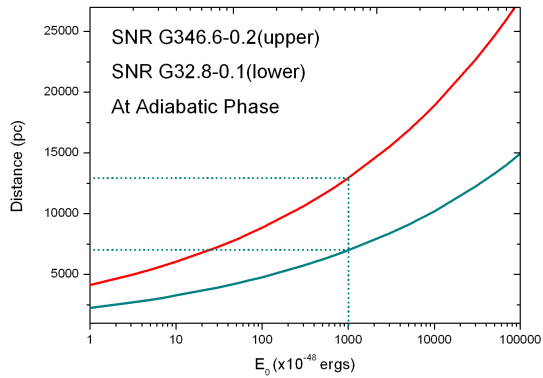}
\caption{The initial explosion
energy ($E_0$) verses the remnant distance ($d$) at Sedov-phase for
both remnants through solving the equations group listed in
Sect.~\ref{sedov}. As the initial energy enhances, the remnants
distance also correspondingly increases. Distance value of both SNRs
is shown when typically $E_0 = 10^{51}$~ergs.}\label{edab}
\end{figure}

We have already known the flux $S_{1GHz} = 11$~Jy at 1~GHz, the
viewing angle $\theta = 17$~arcmin for SNR G32.8$-$0.1, and flux
$S_{1GHz} = 8$~Jy, angle $\theta = 8$~arcmin for SNR G346.6$-$0.2
\citep{gre09} of which we regard both the remnants evolving at the
Sedov-phase. But the linear diameter ($D$) (and then distance
($d$)), evolving age ($t$) and the electron density ($n$) are
unknown and to be derived. After assuming some SNe initial explosion
energy values ($E_0 = 5\times 10^{49}, 10^{50}, 5\times 10^{50},
10^{51}, 5\times 10^{51}$ and $10^{52}$~ergs), then a series of the
diameters $D_{pc}$ (and distance $d_{pc}$) of both SNRs are obtained
by solving the equations group above (table~\ref{tabsedov}).
Furthermore, when the assumed initial explosion energy ($E_0$)
changes from $10^{48}$~ergs to $10^{53}$~ergs, then the kinematic
distance of SNR G32.8$-$0.1 and G346.6$-$0.2 also increases
(Fig.~\ref{edab}). We can obtain a certain distance value for both
remnants corresponding to the typical well-known explosion energy of
the SNRs $E_0 = 10^{51}$~ergs.

\begin{figure}[htp!!]
\epsscale{1.1} \plotone{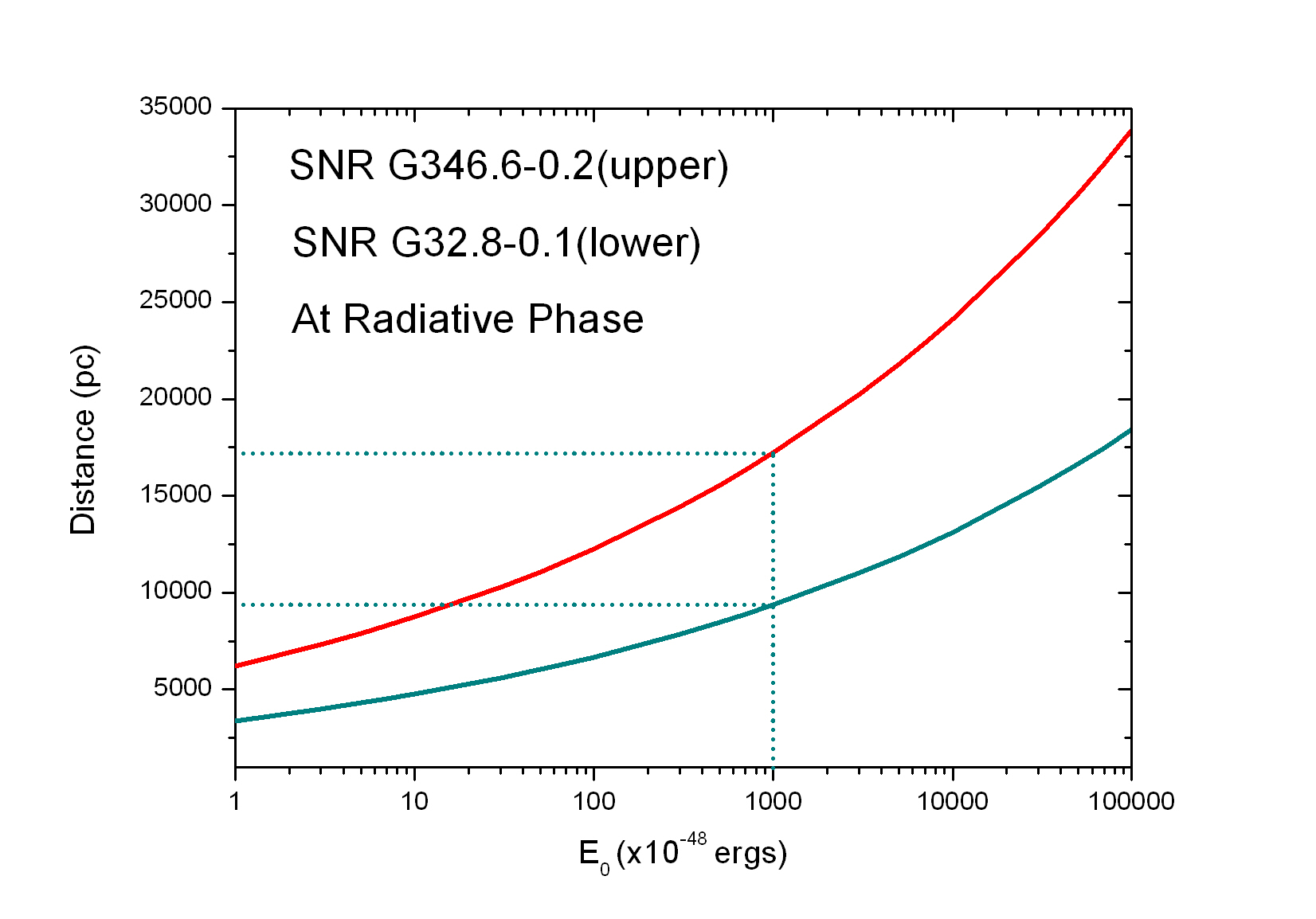}
\caption{The same as in
Fig.~\ref{edab} but at the snowplough-phase.}\label{edcd}
\end{figure}

The group of equations are not strictly correct as not to be figured
out mathematically, but they are correct enough for us to determine
the distance to both SNRs.

From Fig.~\ref{edab} and table~\ref{tabsedov} we can see that the
most likely kinematic distance to SNR G32.8$-$0.1 is about 7~kpc
relevant typically to $E_0 = 10^{51}$~ergs. Since the remnant
diameter evolving at the Sedov-phase is typically less than 36~pc
\citep{cla76,all83,all85}, one can reasonably exclude the two cases
of $E_0 = 10^{52}$~ergs, and $E_0 = 5\times 10^{51}$~ergs for their
too large diameters 50~pc and 45~pc. For the smaller initial energy
is $E_0 = 10^{50}$~ergs and $E_0 = 5\times 10^{49}$~ergs, both their
outcomes are somewhat unlikely. Moreover, the typical SNe initial
explosion energy is $\sim 10^{51}$~ergs as the black numbers show
(table~\ref{tabsedov}). Therefore we subsequently conclude that the
distance to SNR G32.8$-$0.1 is near 7~kpc, that is a little larger
or less than 7~kpc. We can see from next subsection that it is lager
than 7~kpc. The farther distance 8.8~kpc to the remnant is confirmed
as we know later in our work at Sect.~\ref{radiative}.

Similarly the far distance 11~kpc to SNR G346.6$-$0.2 is determined
which is also consistency with the results done in the next
subsection.

\subsection{At radiative phase}\label{radiative}

Analogously we can have the following group of equations for
shell-type remnants at the third stage \citep{koy87,kit05,xu05}

\begin{table}
\begin{center}
\caption{Derived distances and diameters of SNR G32.8$-$0.1 and
G346.6$-$0.2 at Snowplough-phase after assuming several different
initial explosion energies.\label{tabradiative}}
\begin{tabular}{rrrrr}
\tableline \tableline
 & \multicolumn{2}{c}{G32.8$-$0.1}
 & \multicolumn{2}{c}{G346.6$-$0.2}\\
\tableline
 $E_0$(ergs) & d(kpc) & D(pc) & d(kpc) & D(pc)\\
\tableline
 $10^{52}$ & 13 & 65 & 24 & 56\\
 $5\times 10^{51}$ & 12 & 59 & 21 & 51\\
 $\bf{10^{51}}$ & $\bf{9.4}$ & 46 & 17 & 40\\
 $5\times 10^{50}$ & 8.5 & 42 & 16 & 36\\
 $10^{50}$ & 6.7 & 33 & 12 & 29\\
 $5\times 10^{49}$ & 6.0 & 30 & 11 & 26\\
\tableline
\end{tabular}
\end{center}
\end{table}

\begin{equation}
D_{pc} = 1.42 \left( \begin{array}{c}
\frac{E_0/10^{51}ergs}{n_{cm^{-3}}}
\end{array} \right) ^{5/21} t_{yr} ^{2/7}
\end{equation}

\begin{eqnarray}
\Sigma(D) = 1.505\times 10^{-19} \frac{S_{1GHz}}{\theta_{arcmin}^2}\nonumber\\
= 2.88\times 10^{-14} D_{pc}^{-3.8} n_{cm^{-3}}^2
\end{eqnarray}

\begin{equation}
t_{yr} = 10^5 n_{cm^{-3}}^{-3/4} \left( \begin{array}{c}
\frac{E_0}{10^{51}ergs} \end{array} \right)^{1/8}
\end{equation}

Here, $D_{pc}$, $t_{yr}$, $n_{cm^{-3}}$, $S_{1GHz}$ and
$\theta_{arcmin}$ is defined as in Sect.~\ref{sedov} as well as
their units, and $tan \left(\begin{array}{c}
\frac{\theta_{arcmin}}{2} \end{array} \right) =
\frac{D_{pc}}{2d_{pc}}$. The fluxes $S_{1GHz}$ and observational
angle $\theta_{arcmin}$ are already known to us for remnants
G32.8$-$0.1 and G346.6$-$0.2 of which we suppose here both the
remnants evolving at snow-plough phase. When a series of the initial
energies $E_0 = 5\times 10^{49}, 10^{50}, 5\times 10^{50}, 10^{51},
5\times 10^{51}$ and $10^{52}$~ergs are assumed, the remnant
diameters $D_{pc}$ (and distance $d_{pc}$) can be derived by solving
the equations group above (table~\ref{tabradiative}). The same as in
Sect.~\ref{sedov} when the assumed initial energy ($E_0$) enhances
from $10^{48}$~ergs to $10^{53}$~ergs, the SNR G32.8$-$0.1 and
G346.6$-$0.2 distance values also increase (Fig.~\ref{edcd}).
Corresponding to the SNRs typical explosion energy $E_0 =
10^{51}$~ergs one can obtain a certain distance value for both
remnants.

One can see the equations (4) and (6) are rather different from
equations (1) and (3). But formulae (5) and (2) are completely the
same.

From Fig.~\ref{edcd} and table~\ref{tabradiative} we can see that
the most likely distance to SNR G32.8$-$0.1 is about 9.4~kpc
relevant to $E_0 = 10^{51}$~ergs. Therefore the farther distance
8.8~kpc to the remnant is derived when the distance ranges from
7~kpc to 9.4~kpc by combining the results at both evolving stages of
the remnant. One can exclude the two cases of $E_0 = 10^{52}$~ergs
and $E_0 = 5\times 10^{51}$~ergs for their too large distance value
13~kpc and 12~kpc. Other two cases of $E_0 = 5\times 10^{49}$~ergs
and $10^{50}$~ergs can also be removed since their linear diameter
is less than 36~pc which denotes this remnant is not evolving at the
radiative-phase.

\begin{table}
\begin{center}
\caption{Illumination about how large the dispersion of the remnant
kinematic distance would be induced by the difference of remnant
diameters with almost the same surface brightness.
\label{tab:dispe}}
\begin{tabular}{lcc}
\tableline \tableline
 & G32.8$-$0.1 & G346.6$-$0.2\\
\tableline
 $D$(pc)& d(kpc) & d(kpc)\\
\tableline
 18 & 1.8 & 3.9\\
 64 & 6.5 & 13.8\\
\tableline
\end{tabular}
\end{center}
\end{table}

As for the remnant G346.6$-$0.2, in the four casea of $E_0 =
10^{52}$~ergs, $5\times 10^{51}$~ergs, $10^{51}$~ergs and $5\times
10^{50}$~ergs we obtain too large distance values 24~kpc, 21~kpc,
17~kpc and 16~kpc, and could be eliminated. The other two cases $E_0
= 5\times 10^{49}$~ergs and $10^{50}$~ergs are also impracticable
for their small linear diameters 26~pc and 29~pc which means the SNR
is not evolving at the radiative-phase. Therefore SNR G346.6$-$0.2
is most likely evolving at the Sedov-phase with its farther distance
11~kpc which is most near the derived value 13~kpc as its initial
energy equals typically to $10^{51}$~ergs (Fig.~\ref{edab},
table~\ref{tabsedov}).

\section{Discussion}\label{discuss}

\subsection{Why do not use the $\Sigma$-$D$ relation to determine
the distance?}

In literatures many authors make use of the relation between the
radio surface brightness ($\Sigma$) and linear diameter ($D$) of
supernova remnants to estimate the remnant distance. But here we
only adopt a group of equations to confirm the distance to both
remnants, because the $\Sigma$-$D$ relation will lead to very great
deviation on getting a remnant distance.

One can find that the statistical fitting line in the $\Sigma$-$D$
plot by \citep{cas98},
\begin{eqnarray}
\Sigma(D) = 5.43\times 10^{-17} D_{pc}^{-2.64}\nonumber\\
~(W m^{-2} Hz^{-1} sr^{-1})
\end{eqnarray}
was widely used to determine the remnant distance when other better
methods are not available. As we analyze the data (table 1 in
\citet{cas98}) and plot (Fig. 1 in \citet{cas98}) one can discover
that in some case the dispersion of the derived diameter (then the
distance) for a certain remnant could be quite large. For example,
at the diameter $D = 18$~pc, the SNR G327.6+14.6 has the radio
surface brightness $\Sigma = 3.2\times 10^{-21}$~$W m^{-2} Hz^{-1}
sr^{-1}$, and at $D = 64$~pc, the SNR G359.1$-$0.5 has nearly the
same brightness $\Sigma = 3.7\times 10^{-21}$~$W m^{-2} Hz^{-1}
sr^{-1}$. The relevant kinematic distance for SNR G32.8$-$0.1 with
the viewing angle $\theta = 17$~arcmin would be one 1.8~kpc and
another 6.5~kpc (table~\ref{tab:dispe}). For SNR G346.6$-$0.2 with
viewing angle $\theta = 8$~arcmin the corresponding distance could
be one 3.9~kpc and another 13.8~kpc. It is obvious that such great
dispersion on the obtained distance is truly impracticable.
Furthermore, if the SNR G359.1$-$0.5 continuously evolves on until
the surface brightness reaches the same value $\Sigma = 3.2\times
10^{-21}$~$W m^{-2} Hz^{-1} sr^{-1}$ as that for SNR G327.6+14.6,
then the remnant diameter could be larger than 64~pc, which denotes
much bigger dispersion for the derived kinematic distance. There are
many other different statistical results for the fitting line of the
remnants $\Sigma$-$D$ relation
\citep{pov68,loz81,hua85,dur86,gus03}, of which their data
dispersion shows much larger. Thus they will lead to bigger
deviation when being used to determine the SNRs distance. Therefore
we use the group of equations listed in the text to determine the
SNR distance can avoid such large dispersion and reach a somewhat
good outcome.

\subsection{Other remarks}

It seems that the SNR G32.8$-$0.1 may be at the phase-transformation
stage just evolving from the Sedov-phase to the snowplough-phase
with its linear diameter about 36~pc (table~\ref{tabsedov}) roughly
at which the SNRs transform from one phase to another. But the SNR
G346.6$-$0.2 may probably be still at the second-phase. Some days in
the future a much better method to estimate this two remnants
distance may be available, and the same as a more delicate detect
technique which would test or proof our result.

\acknowledgments

JWX like to thank H.B. Zhang and Y.P. Hu et al. for helpful
discussion.

\end{document}